\documentclass[10pt, twocolumn, nofootinbib,superscriptaddress]{revtex4-1}
\usepackage{amsmath,amssymb,amsfonts}
\usepackage{algorithmic}
\usepackage{graphicx}
\usepackage{textcomp}
\usepackage{xcolor}
\usepackage{ragged2e}
\usepackage{booktabs, makecell, tabularx}

\usepackage{gensymb}
\makeatletter
    \renewcommand\@make@capt@title[2]{%
     \@ifx@empty\float@link{\@firstofone}{\expandafter\href\expandafter{\float@link}}%
      {\textbf{#1}}\@caption@fignum@sep#2\quad}%
\makeatother
\makeatletter 
\renewcommand{\fnum@figure}{\textbf{Fig.~\thefigure}}
\makeatother
\usepackage{svg}

\usepackage{xcolor}
\newcommand{\beginsupplement}{%
        \setcounter{table}{0}
        \renewcommand{\thetable}{S\arabic{table}}%
        \setcounter{figure}{0}
        \renewcommand{\thefigure}{S\arabic{figure}}%
        \setcounter{equation}{0}
        \renewcommand{\theequation}{S\arabic{equation}}%
     }
\def\BibTeX{{\rm B\kern-.05em{\sc i\kern-.025em b}\kern-.08em
    T\kern-.1667em\lower.7ex\hbox{E}\kern-.125emX}}
\begin{document}
\title{Stimulated Brillouin scattering in tellurite-covered silicon nitride waveguides}

\author{Roel~A.~Botter\textsuperscript{\textdagger}}
\author{Yvan~Klaver\textsuperscript{\textdagger}}
\author{Randy~te~Morsche\textsuperscript{\textdagger}}
\affiliation{Nonlinear Nanophotonics, MESA+ Institute of Nanotechnology, University of Twente, Enschede, the Netherlands}
\author{Bruno~L.~Segat~Frare}
\author{Batoul~Hashemi}
\affiliation{Bradley Research Group, Department of Engineering Physics, McMaster University, Hamilton, Ontario, Canada}
\author{Kaixuan~Ye}
\author{Akhileshwar~Mishra}
\author{Redlef~B.G.~Braamhaar}
\affiliation{Nonlinear Nanophotonics, MESA+ Institute of Nanotechnology, University of Twente, Enschede, the Netherlands}
\author{Jonathan~D.B.~Bradley}
\affiliation{Bradley Research Group, Department of Engineering Physics, McMaster University, Hamilton, Ontario, Canada}
\author{David~Marpaung}
\email{Corresponding author: david.marpaung@utwente.nl}
\affiliation{Nonlinear Nanophotonics, MESA+ Institute of Nanotechnology, University of Twente, Enschede, the Netherlands}

\begin{abstract}


Stimulated Brillouin scattering (SBS), a coherent nonlinear effect coupling acoustics and optics, can be used in a wide range of applications such as Brillouin lasers and tunable narrowband RF filtering. Wide adoption of such technologies however, would need a balance of strong Brillouin interaction and low optical loss in a structure compatible with large scale fabrication. Achieving these characteristics in scalable platforms such as silicon and silicon nitride remains a challenge. Here, we investigate a scalable Brillouin platform combining low loss Si$_3$N$_4$ and tellurium oxide (TeO$_2$) exhibiting strong Brillouin response and enhanced acoustic confinement. In this platform we measure a Brillouin gain coefficient of 8.5~m$^{-1}$W$^{-1}$, exhibiting a twenty fold improvement over the largest previously reported Brillouin gain in a Si$_3$N$_4$ platform. Further, we demonstrate cladding engineering to control the strength of the Brillouin interaction. We utilized the Brillouin gain and loss resonances in this waveguide for an RF photonic filter with more than 15 dB rejection and 250 MHz linewidth. Finally, we present a pathway by geometric optimization and cladding engineering to a further enhancement of the gain coefficient to 155~m$^{-1}$W$^{-1}$, a potential 400 times increase in the Brillouin gain coefficient.

\end{abstract}

\maketitle
\def\thefootnote{\textdagger}\footnotetext{These authors contributed equally to this work}\def\thefootnote{\arabic{footnote}}

\section*{Introduction}

Stimulated Brillouin scattering (SBS), a nonlinear optical effect mediated through acoustic waves is a burgeoning field, which can be used in a wide range of fields, from telecommunications to sensing \cite{Safavi-Naeini_Optica_2019, Eggleton_NatPhot_2019, Rakich_PRX_2012}. SBS results in a narrowband (tens of MHz) gain resonance, shifted from the pump frequency by about 10 GHz. This makes it a unique filter and amplifier with applications in next-generation optical and radio communications \cite{Marpaung_NatPhot_2019, Gertler_APLPhot_2020}, low-threshold narrow-linewidth lasers \cite{Kabakova_OptLett_2013, Gundavarapu_NatPhot_2018}, nonreciprocal light propagation \cite{Kittlaus_NatPhot_2018, Sohn_NatPhot_2018} and high-precision sensors \cite{Denisov_LightSciAppl_2016, Lai_NatPhot_2020}.

To achieve wide adoption of SBS in applications, the Brillouin nonlinearity needs to be integrated in scalable photonic platforms. Traditionally, high on-chip Brillouin gain in standard photonic platforms has been achieved by enhancing the effective length through low-loss waveguides \cite{Gundavarapu_NatPhot_2018}, or by increasing the Brillouin gain coefficient. The latter can be achieved by unlocking acoustic guiding via the waveguide geometry \cite{VanLaer_NJP_2015, Kittlaus_NatPhot_2016} or by adding waveguides made of high-gain materials through hybrid \cite{Morrison_Optica_2017, Lai_AFM_2022, Kim_ncomms_2020} or heterogeneous integration \cite{Garrett_IPC_2022}. However, these platforms come with several challenges, such as low integration density, less robust free-floating structures, multi-photon absorption, or a lack of tuneable components. To circumvent these challenges, currently there are efforts to achieve improved acoustic waveguiding in nitride based platforms such as silicon nitride \cite{Botter_SciAdv_2022} and silicon oxynitride \cite{Ye_APL_2023, Zerbib_ArXiv_2023}. However, the acoustic guidance is fundamentally limited by the lack of confinement of the acoustic wave within the nitride itself due to the hardness of the core. 

\begin{figure*}
    \centering
    \includegraphics[width=\textwidth]{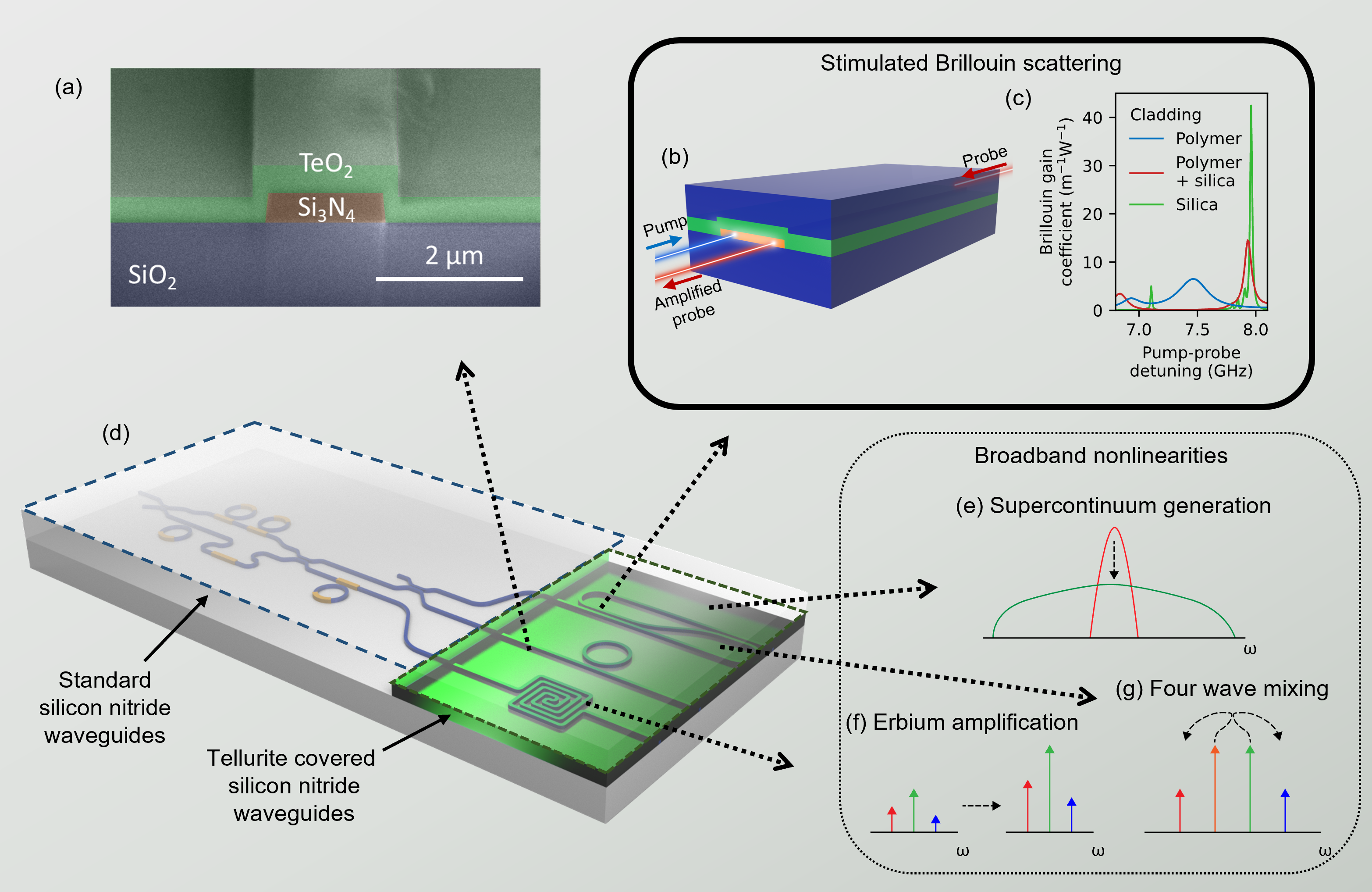}
    \caption{\textbf{Combining nonlinear tellurite covered waveguides with a scalable silicon nitride photonic integrated circuit.} (a) A false colour SEM image of an (uncladded) tellurite covered silicon nitride waveguide. (b) Stimulated Brillouin scattering in tellurite covered waveguides, as described in this work. (d) Our vision of a device combining standard silicon nitride with tellurite covered silicon nitride. The tellurite covered waveguides also support other nonlinear processes, including (e) supercontinuum generation, (f) on-chip amplification through erbium doping and (g) four wave mixing.}
    \label{fig:fig1}
\end{figure*}

Recently, a hybrid integration platform that combines standard wafer-scale thin silicon nitride and tellurium oxide (TeO$_2$, tellurite) has been investigated, see Fig.~\ref{fig:fig1}~(a) and (d). These tellurite-covered silicon nitride waveguides create a promising platform for low loss circuits, with losses down to 0.25~dB/cm \cite{Frankis_OptLett_2018, Frankis_OptExp_2019}, by adding a single CMOS compatible back-end step \cite{Singh_PhotRes_2020}. Tellurite itself is known to be a good platform for acousto-optic devices\cite{Yano_JApplPhys_1971}, where SBS has been previously demonstrated in tellurite fibers \cite{Abedin_OE_2006} and observed in integrated waveguides \cite{Morrison_Thesis_2017}. The tellurite-covered silicon nitride waveguides platform has shown a variety of applications in on-chip amplifiers \cite{Frankis_PhotRes_2020}, four-wave mixing \cite{Kiani_OSACont_2020}, octave-spanning supercontinuum and third-harmonic generation \cite{Singh_PhotRes_2020, Mbonde_FIO_2022}, as depicted in Fig.~\ref{fig:fig1}~(e) through (g). Although highly promising, the Brillouin response of these waveguides, illustrated in Fig.~\ref{fig:fig1}~(b)~and~(c), had not yet been investigated. 

In this work, we present a comprehensive study of SBS in tellurite-covered silicon nitride waveguides. We experimentally demonstrate, for the first time, SBS in such hybrid waveguides. We study the impact of cladding engineering to tailor the strength of Brillouin interactions in these waveguides. We further demonstrate RF photonic filtering through harnessing the Brillouin gain and loss resonances in our sample. Finally, we perform a geometric optimization via simulations, showing the possibility to improve the Brillouin gain factor by two orders of magnitude. These results represent the highest gain coefficient achieved in a silicon nitride-based circuit and open the possibility towards integrating a Brillouin engine in a tunable complex circuit for applications such as narrow linewidth lasers, frequency combs, and RF photonic signal processors.

\section*{Results}

\subsection*{SBS in tellurite-covered waveguides}

\begin{figure*}
    \centering
    \includegraphics[width=\linewidth]{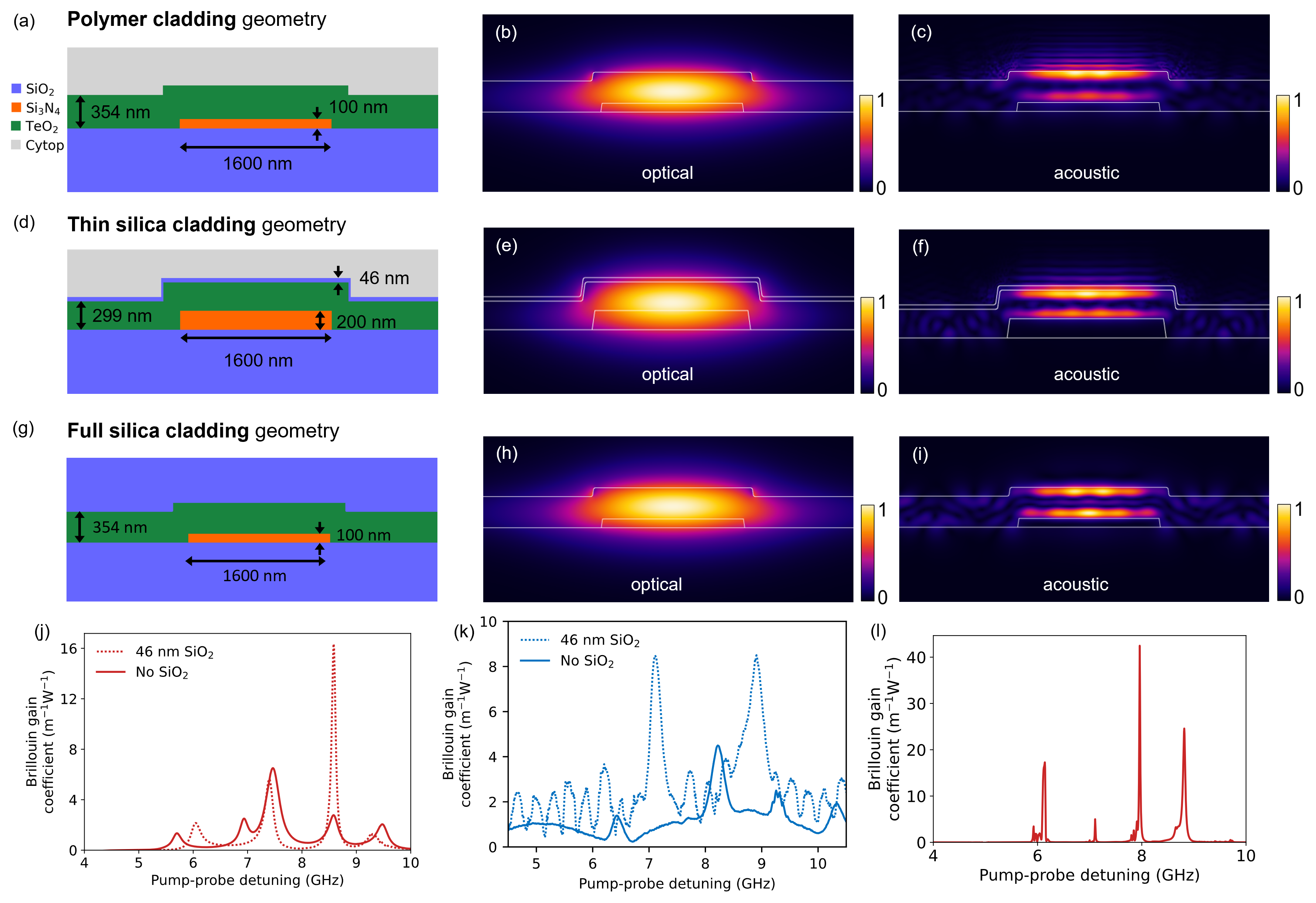}
    \caption{\textbf{Cladding engineering for enhanced Brillouin scattering.} We investigetd three different waveguide geometries; with a CYTOP cladding, with a 46~nm silicon oxide barrier added, and with a full silicon oxide cladding. (a, d, g) The geometry of the waveguides, with and without silicon oxide barrier. (b, e, h) The simulated electric field of the optical modes, and (c, f, i) the simulated displacement fields of the acoustic response at the acoustic frequency with the highest Brillouin gain. For the CYTOP covered waveguides with and without silicon oxide barrier, shown are (j) the simulated Brillouin responses compared to (k) the measured Brillouin responses. (l) Simulated Brillouin response of the full silicon oxide cladded waveguide.}
    \label{fig:fig2}
\end{figure*}

Our samples consist of single-stripe  optical waveguides in LPCVD silicon nitride  with a height of 100~nm, and width of 1600~nm. The entire chip is then covered with a layer of tellurite with a height of 354~nm (see Methods for details of fabrication). Fig.~\ref{fig:fig2}~(a) shows the geometry of the waveguide. The waveguides are cladded with a layer of CYTOP, a fluoropolymer, to protect the waveguiding layer.

We investigate the strength of Brillouin scattering in these waveguides through simulations of the optical and acoustic modes, implemented in COMSOL Multiphysics. We start with an optical simulation as depicted in Fig.~\ref{fig:fig2}~(b). This optical simulation can be used to calculate the optical forces, which we use as a basis for the calculation of the acoustic response. The acoustic field at the highest Brillouin peak is shown in Fig.~\ref{fig:fig2}~(c). The overlap between the optical and acoustic fields is used to calculate the Brillouin interaction strength, and create a response spectrum, which is depicted as the solid line in Fig.~\ref{fig:fig2}~(j). The details of the simulation methods can be found in the Supplementary Materials. The simulations revealed the signature of SBS with a gain coefficient of 6.5~m$^{-1}$W$^{-1}$ at a frequency shift of 7.47~GHz.

We proceeded with the measurements of the Brillouin gain response of a 1~cm long waveguide. We measured a propagation loss of 1~dB/cm in the sample. We then implemented a double intensity modulation pump-probe scheme \cite{Botter_SciAdv_2022} for enhanced-sensitivity SBS characterisations. The details of the measurement apparatus are described in the Methods section and Supplementary materials. The measured Brillouin response can be seen in Fig.~\ref{fig:fig2}~(k) with the highest Brillouin peak gain of 4.5~m$^{-1}$W$^{-1}$ appearing at 8.2~GHz Stokes-shifted frequency from the pump. This gain coefficient is an order of magnitude higher than that of silicon nitride waveguides reported recently \cite{Botter_SciAdv_2022, Gundavarapu_NatPhot_2018, Gyger_PRL_2020}. The measurement and simulation results show good agreement in both the magnitude and frequency shift of the SBS gain peak.

\subsection*{SBS enhancement through cladding engineering}

\begin{table*} [th!]
\centering
    \caption{\textbf{SBS characteristics of cladding-engineered waveguides.}}
\begin{tabular}{c|c|c|c|c}
\textbf{Geometry} & \multicolumn{2}{c|}{\textbf{Simulated}} & \multicolumn{2}{c}{\textbf{Measured}}               \\
 & \textbf{max} $\mathbf{g_B}$ & \textbf{SBS shift} & \textbf{max} $\mathbf{g_B}$ & \textbf{SBS shift} \\  
 & (m$^{-1}$W$^{-1}$) & (GHz) & (m$^{-1}$W$^{-1}$) & (GHz) \\
                           \hline 
Polymer cladding                & 6.5   & 7.47  & 4.5  &  8.2             \\
Thin silica cladding            & 16.4  & 8.58  & 8.5  &  8.9             \\
Full silica cladding            & 42.5  & 7.96  & -    & -                \\
Optimized full silica cladding  & 154.8   & 6.335 & -    & -              
\end{tabular}
\label{tab:gain_summary}
\end{table*}

To further improve the Brillouin gain of these waveguides we analyze their acoustic behaviour. As seen in Fig.~\ref{fig:fig2}~(c),  significant acoustic wave leakage occurs from the tellurite core to the CYTOP polymer upper cladding due to lack of contrast in acoustic impedance. This reduces the strength of the acoustic waves in the waveguide, thereby reducing the Brillouin gain. Furthermore, the polymer is acoustically lossy, attenuating the acoustic fields, and thus further decreasing the Brillouin gain. Here we circumvent these adverse effects through cladding engineering that provides improved acoustic waveguiding without degrading the optical propagation losses. 

 A good candidate for cladding material to prevent acoustic acoustic leakage is silicon oxide. Ideally, a thick (over a micron) cladding will allow for optimized acoustic waveguiding (see Fig.~\ref{fig:fig2}~(i), for example). However, current limitations in our fabrication process only allow low-temperature deposition (required to avoid increased losses in the tellurite layer \cite{Nayak_TSL_2002}) of low-loss silica of the order of a few tens of nanometer. We therefore investigated the effectivenes of using a thin layer of silicon oxide to reduce acoustic leakage into the polymer and to increase the Brillouin gain.

The simulated SBS response of a thin (46~nm) silica cladded waveguide is summarized in Fig.~\ref{fig:fig2}~(d)-(f) and (j). Appreciable reduction of the acoustic wave amplitude leaking in to the upper polymer cladding was observed. The calculated Brillouin gain coefficient of this waveguide is 16.4~m$^{-1}$W$^{-1}$, which is more than double that of the gain without the thin silica cladding. This result shows the effective impact of cladding engineering even using a very thin cladding layer.    

We then fabricated the cladding-engineered waveguide, and measured the propagation loss and the SBS gain coefficient. The waveguide consists of a 200~nm layer of silicon nitride, covered with a 299~nm layer of tellurite. The cladding consists of a 46~nm layer of silicon oxide, and again a layer of polymer as protection. We performed SBS characterisations in a 10~cm long waveguide with loss of  0.59~dB/cm. The measured gain response is shown in Fig.~\ref{fig:fig2}~(k) and shows two peaks of equal magnitude, each with a Brillouin gain coefficient of 8.5~m$^{-1}$W$^{-1}$. The gain coefficient is double of that of the waveguides with only polymer cladding, demonstrating the effectiveness of our cladding engineering strategy. 

As mentioned earlier, optimum performance can be unlocked using thick a silica cladding. We simulated a waveguide geometry with a similar cross-section as the polymer-cladded waveguide considered earlier, only that the cladding is replaced by microns-thick silicon oxide to provide full acoustic waveguiding. The detailed waveguide geometry and acousto-optic characteristics are shown in Fig.~\ref{fig:fig2}~(d)-(f). As a result of improved acousto-optic overlap, enhancement of the the Brillouin gain coefficient beyond 40~m$^{-1}$W$^{-1}$ can be achieved, as shown in Fig.~\ref{fig:fig2}~(l). The simulated and measured SBS gain of the cladding engineered waveguides are summarized in Table~\ref{tab:gain_summary}.

\subsection*{RF photonic notch filter} 

\begin{figure*} [t]
    \centering
    \includegraphics[width=\linewidth]{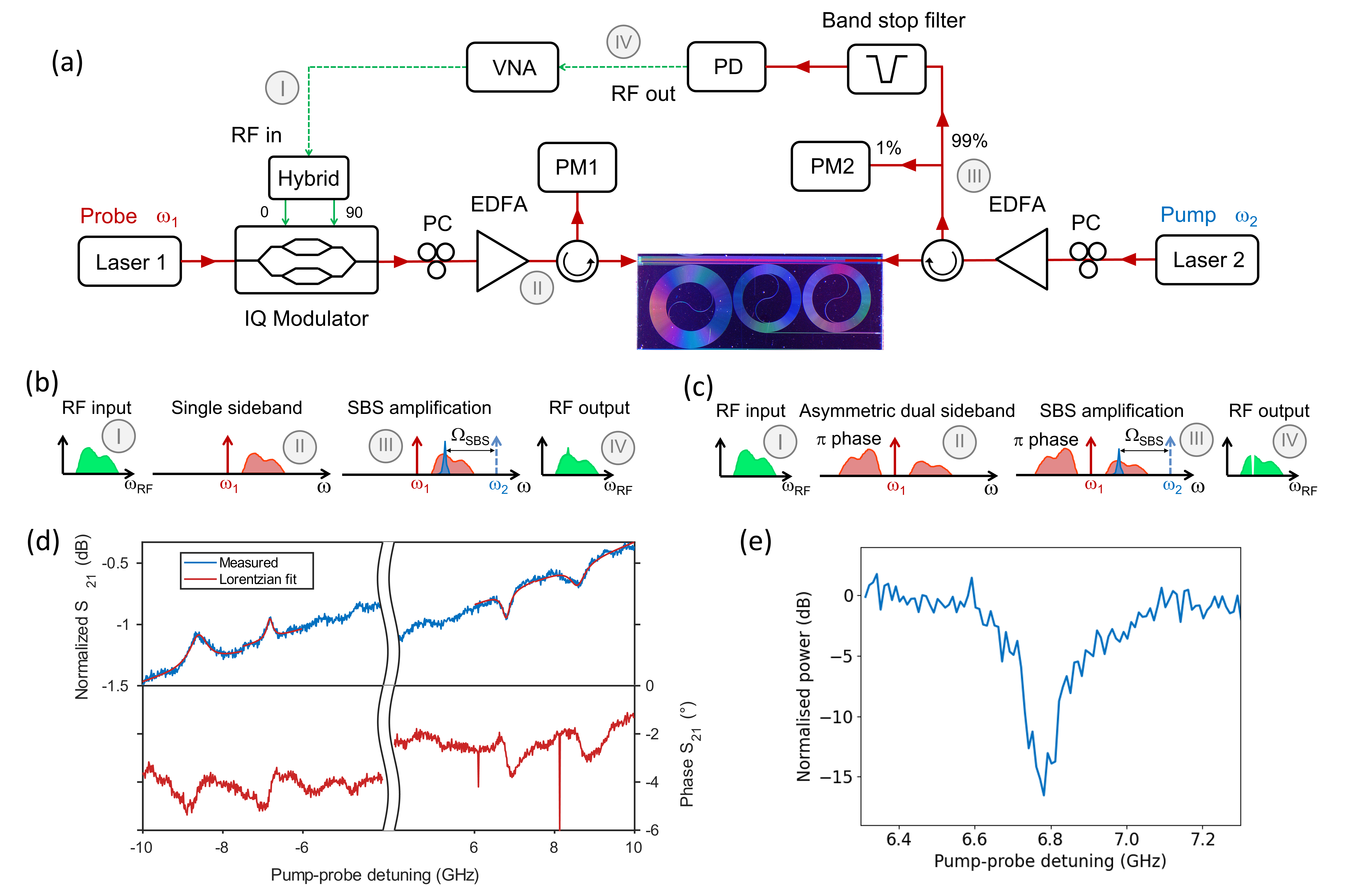}
    \caption{\textbf{RF photonic filter in tellurite-covered silicon nitride waveguide.} (a) Schematic of the measurement setup of the Vector Network Analyzer (VNA) based measurements. IQ mod: In-phase and Quadrature modulator, PD: photodiode, PM: Power meter. (b) RF and optical spectra at different stages of the SBS gain measurement. (c) RF and optical spectra at different stages of the RF filter. (d) $S_{21}$ gain and phase of VNA based SBS gain measurement results. (e) The notch filter realised using the on-chip SBS gain, showting enhanced 15~dB rejection from only 0.2~dB of SBS gain.}
    \label{fig:fig3}
\end{figure*}

We demonstrate the use of the on-chip Brillouin gain in our cladding-engineered tellurite-covered waveguide in a cancellation-based RF photonic notch filter experiment \cite{Marpaung_Optica_2015}. In such a filter, relatively low SBS gain or loss can be tranformed to a high extiction notch filter through RF destructive interference. In this case, a modulator is used to prepare an RF modulated spectrum consisting of an optical carrier and two RF sidebands with opposite phase and unbalanced amplitude, before one of these sidebands is processed by the SBS gain of loss resonances \cite{Marpaung_Optica_2015}.

We devise a measurement setup based on a vector network analyzer (VNA) to accurately measure the on-off SBS gain and loss from the sample, as illustrated in Fig.~\ref{fig:fig3}~(a). The synthesized single-sideband with carrier optical signal from the IQ modulator was injected to a 15~cm long spiral waveguide that is pumped with 26.6~dBm of on-chip power at 1561 nm.  We then detect the amplification and loss of the sideband at the SBS shift frequencies using a photodetector. The signal flow of such a characterization step is illustrated in Fig.~\ref{fig:fig3}~(b). 

In Fig.~\ref{fig:fig3}~(f) we show the resulting $S_{21}$ relative to a pump-off measurement of the VNA based measurement that indicates a total amplification for the Stokes signal of 0.2~dB for both peaks with a full width half maximum (FWHM) of 500 MHz and 200 MHz at 8.6 GHz and 6.8 GHz respectively, which is identical for the absorption observed at the anti-Stokes frequencies.

To achieve an RF photonic notch filter from the gain resonance, the IQ modulator is set to generate a dual sideband (DSB) modulated signal with opposite phases, so there is a $\pi$ phase difference between the lower sideband (LSB) and upper sideband (USB). The power imbalance between the LSB and USB is set equal to the Brillouin gain by tuning the DPMZM bias voltages. By now adding a pump laser, appropriately detuned from the signal carrier (probe) laser, signal cancellation is achieved within the Brillouin gain window. This is shown schematically in Fig.~\ref{fig:fig3} (c). The normalized response of the measured RF photonic notch filter is shown in Fig.~\ref{fig:fig3}~(e). Due to the limited ($\sim$ 0.2~dB) gain in the waveguide, only a limited rejection of about 15~dB could be realised. Further improvements in the technology, including SBS gain enhancement through cladding engineering, interfacing with an on-chip pre-processing tunable circuit like ring resonators \cite{Botter_SciAdv_2022}, and the possibility to achieve on-chip amplification through erbium doping \cite{Frankis_OptExp_2019} promise an encouraging route to the realization of a high performance SBS RF photonic filter in this platform.

\subsection*{Enhancing the Brillouin response by geometric optimization}
\begin{figure*}
    \centering
    \includegraphics[width=\linewidth]{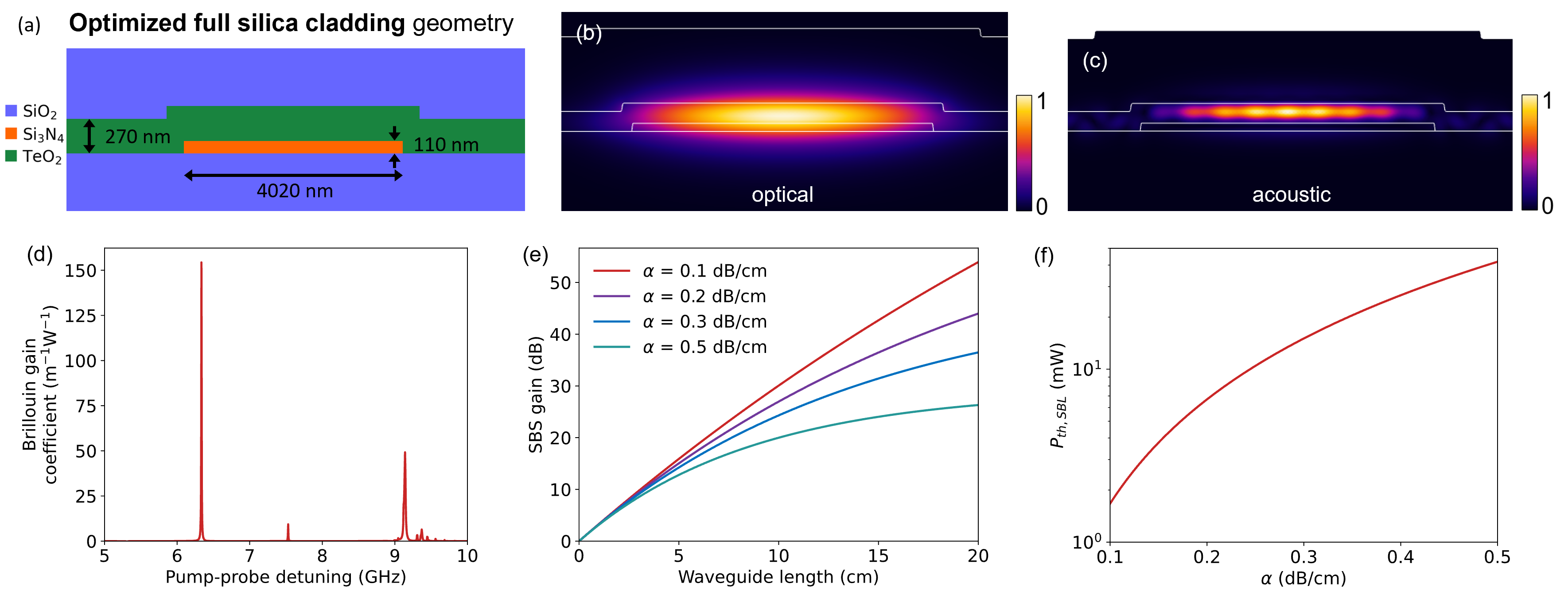}
    \caption{\textbf{Gain optimization of tellurite-covered silicon nitride waveguide.} (a) optimized geometry of the waveguides with 1~$\mu$m silicon oxide cladding. (b) Simulated electric field of the optical modes, and (c) the simulated displacement fields of the acoustic response at the acoustic frequency with the highest Brillouin gain. (d) Simulated Brillouin gain spectra of the full silicon oxide cladded waveguide exhibiting peak gain coefficient of 155~m$^{-1}$W$^{-1}$. (e) SBS gain as functions of waveguide length and propagation loss with 500~mW of on-chip optical power. (f) Stimulated Brillouin lasing threshold in a single ring resonator, with FSR matched to the Brillouin shift.}
    \label{fig:fig4}
\end{figure*}

Apart from the cladding material, the Brillouin response is also sensitive to the geometry of the waveguide cross-section. In order to obtain the maximal improvement, we performed a genetic optimization on the cross-sectional dimensions of the structure with full silicon oxide cladding.

The geometry with the highest gain is shown in Fig.~\ref{fig:fig4} (a),  with a silicon nitride width and thickness of 4020~nm and 110~nm, respectively, and a tellurite thickness of 270~nm. The cladding thickness was found to have little influence beyond 800~nm, thus it is given a set value of 1~$\mu$m. The optical mode and acoustic response are shown in Fig.~\ref{fig:fig4}~(b) and (c), respectively. The acoustic response shows an acoustic mode with a single vertical lobe which corresponds to the lowest frequency peak in the gain spectrum. Compared to the acoustic responses of the waveguides in Fig.~\ref{fig:fig2}, the optimized waveguide achieves significant gain enhancement through improved vertical acoustic confinement and acousto-optic overlap. The peak Brillouin gain of the optimized waveguide is estimated to be  155~m$^{-1}$W$^{-1}$ with a linewidth of $\Gamma = 5.71$~MHz, as shown in Fig.~\ref{fig:fig4}~(d). The details of of the optimization procedure and more extensive results can be found in Supplementary Note~E.

With the optimized structures the gain is significantly improved, to illustrate the impact we show the calculated gain for different lengths of waveguides in Fig.~\ref{fig:fig4}~(e). Pumped with a power of 27 dBm these show significant gain of 10s of dB for even realistic losses close to those observed. Such numbers come close to what has been achieved in chalcogenide based waveguides \cite{Eggleton_NatPhot_2019}.

This high gain can also be leveraged in the use of on-chip Brillouin lasers \cite{Kabakova_OptLett_2013,Morrison_Optica_2017, Otterstrom2018ALaser, Gundavarapu_NatPhot_2018}, leading to lower thresholds. As an example for the optimized waveguide geometry for a full oxide cladding, the thresholds can be estimated as given in Fig.~\ref{fig:fig4} (f). Here, the length of the cavity is matched to 1 FSR using a group index of 1.635, and assuming a design for minimal threshold calculated as \cite{Botter_SciAdv_2022}:
\begin{equation}\label{eq:threshold}
    P_{\rm{th}} = 0.089\frac{\alpha^2 L}{g_{\rm{B}}},
\end{equation}
where $g_{\rm{B}}$ is the Brillouin gain in m$^-1$W$^-1$, $\alpha$ is the propagation loss in dB/cm and $L$ is the cavity length, that is 2.9~cm in this case. As can be seen in Fig.~\ref{fig:fig4}~(f), even at an expected loss of 0.5~dB/cm the large gain still means a threshold of a few 10s of mW. This would lead to practical pump power levels when applied even for these large diameter ring resonators.

\section*{Discussion}

We have shown, for the first time, the Brillouin response of tellurite covered silicon nitride waveguides. We enhanced the Brillouin gain by adding a thin layer of silicon oxide to reduce the acoustic losses. This resulted in a Brillouin gain coefficient of 8.5~m$^{-1}$W$^{-1}$, a record for silicon nitride based waveguides.

By genetic optimization, we determined an optimal geometry for a tellurite covered silicon nitride waveguide with a full silicon oxide cladding. The improved gain in this waveguide with optimization of the cross-sectional geometry is 155~m$^{-1}$W$^{-1}$. This represents a 400 times increase in the Brillouin gain coefficient compared to previous results in silicon nitride waveguides.  The gain is significantly higher, such that an ultra-narrow amplifier with 10s of dB of gain or a simple design Brillouin laser with a few 10s of mW threshold should be achievable.

Various improvements in the fabrication and design can lead to realizing the vision of such high-Brillouin-gain waveguides in silicon nitride PICs, including adding full hard cladding and reducing optical propagation losses. We also aim toward developing tapers from waveguides without tellurite to tellurite covered waveguides. This will open up the possibility of combining standard and versatile silicon nitride waveguide circuits with high Brillouin gain sections enabled by the tellurium oxide layer, as well as low loss edge-couplers for efficient fiber-chip pumping and signal coupling.. This will represent a novel concept of spatial selection of Brillouin enhancement in a large scale circuit. 

\begin{figure*}
    \centering
    \includegraphics[width=0.9\textwidth]{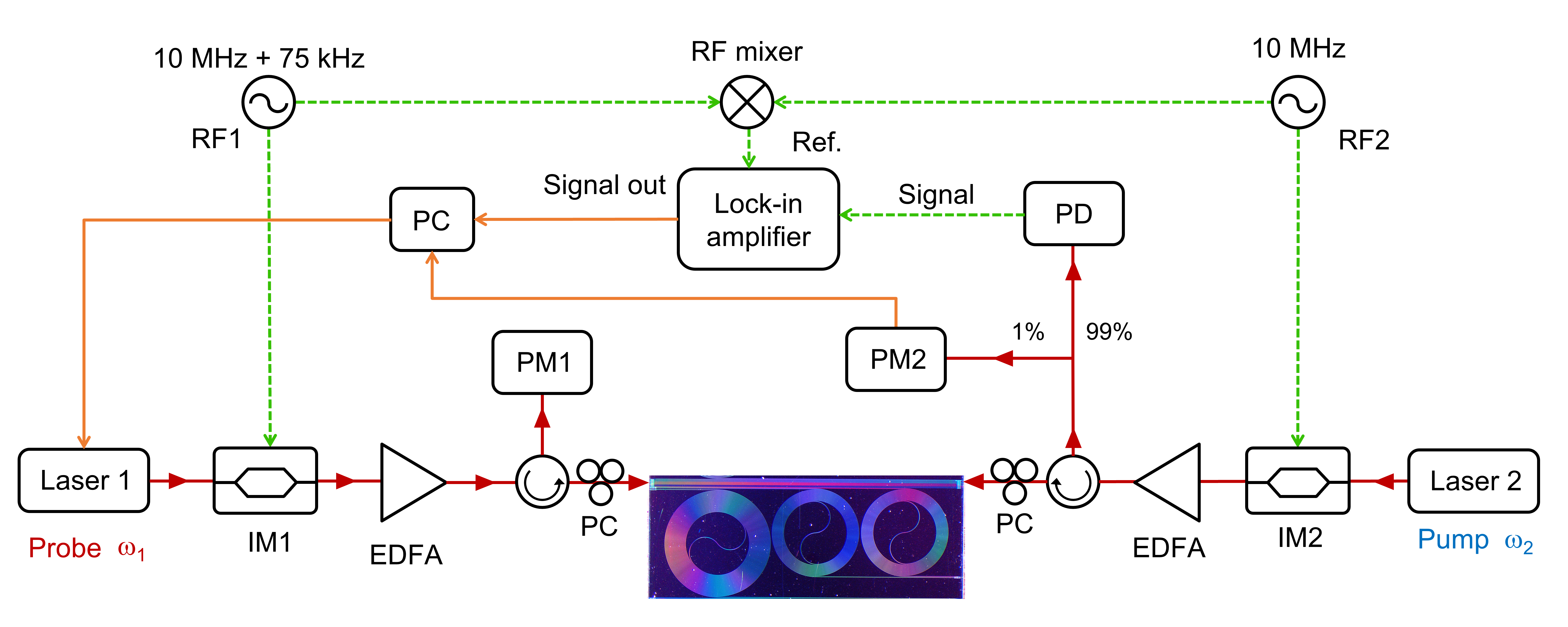}
    \caption{\textbf{Schematic of the measurement setup used for the SBS gain measurements.} IM: intensity modulator, PC: computer, PD: photodiode, PM: power meter.}
    \label{fig:fig5}
\end{figure*}
The high Brillouin gain of these waveguides, combined with the other non-linearities, opens an avenue towards the creation of Brillouin-Kerr soliton frequency combs \cite{Bai_PRL_2021}. These high non-linearities can create these solitons without requiring extremely low-loss waveguides.

\section*{Methods}
\subsection*{Waveguide fabrication}

The silicon nitride strips were fabricated on a 100~mm diameter silicon wafer using a standard foundry process \cite{Roeloffzen_JSTQE_2018}. This involves growing an silicon oxide through wet thermal oxidation of the silicon substrate at 1000°C. The silicon nitride layer was then grown using low-pressure chemical vapor deposition (LPCVD). The waveguides were patterned into the silicon nitride layer by stepper lithography
and reactive ion etching.

The wafer was then diced into individual chips. Next the tellurite layer, and where applicable the silicon oxide layer, was grown using radio frequency (RF) reactive sputtering \cite{Frankis_OptExp_2019}. Finally, the polymer layer was applied via spincoating.

\subsection*{Brillouin gain measurement}

The measurements were performed using an updated version of the double intensity modulation setup previously used in our work on multilayer silicon nitride waveguides \cite{Botter_SciAdv_2022}, which was based on earlier work in Brillouin spectroscopy \cite{Grubbs_RSI_1994} and gain measurements in thick silicon nitride waveguides \cite{Gyger_PRL_2020}. In this setup, schematically illustrated in Fig.~\ref{fig:fig5}, both the pump and probe laser are modulated at slightly different frequencies. This way any interaction between them when traveling through the sample (i.e. SBS) will result in a signal at their difference frequency such that the interaction strength can be recorded using a lock-in amplifier. A detailed description can be found in Supplementary Note~A.

\section*{Author contributions}

D.M. and R.A.B. developed the concept and proposed the physical system. R.A.B., R.M. and K.Y. developed and performed numerical simulations. R.M. performed the numerical optimization with input from K.Y. and Y.K.. R.A.B. performed the gain characterization experiments, with input from B.L.S.F., B.H., K.Y. and Y.K.. Y.K. and R.M. performed the VNA based experiments with input from D.M., A.M. and R.B.G.B.. B.L.S.F., B.H. and J.D.B.B. developed and fabricated the samples. D.M., R.A.B., Y.K. and R.M. wrote the manuscript. D.M. supervised the project.


\section*{Funding Information}

The authors acknowledge funding from the European Research Council Consolidator Grant (101043229 TRIFFIC), Nederlandse Organisatie voor Wetenschappelijk Onderzoek (NWO) Vidi (15702) and Start Up (740.018.021), the Natural Sciences and Engineering Research Council of Canada (NSERC) (I2IPJ 555793-20 and RGPIN-2017-06423) and the Canadian Foundation for Innovation (CFI) (35548).

\bibliographystyle{IEEEtran}
\bibliography{library}

\newpage
\onecolumngrid
\beginsupplement
\newpage

\section*{Supplementary note A: Measurement setups}

As mentioned in the methods of the main text, the measurements were performed using an updated version of the double intensity modulation setup where both the pump and probe laser are modulated at slightly different frequencies. This way the SBS interaction between them will result in a signal at their difference frequency. 

A schematic overview of the setup can be seen in Fig.~\ref{fig:fig5}. The setup uses a probe laser (Toptica DFB pro BFY), operating around 1550~nm, which is scanned using current control. This laser is modulated with a 10.075~MHz sine wave generated by a Hewlett-Packard 33120A function/arbitrary waveform generator (AWG) using a Thorlabs LN82S-FC intensity modulator. This signal is then amplified using an Amonics AEDFA-PA-35 fiber amplifier.

The pump laser is an Avanex A1905LMI, operating at a fixed wavelength around 1550~nm. This light is modulated with a 10~MHz sine wave produced by a Wiltron 69147A synthesized sweep generator (SSG) using another Thorlabs LN82S-FC intensity modulator. The output is then amplified using an Amonics AEDFA-33-B fiber amplifier. 

The SBS signal is converted from to the electrical domain using a Discovery Semiconductor DSC30S photodiode, and analyzed using a Zurich Instruments MFLI lock-in amplifier. The reference signal is made by mixing the 10~MHz reference output of the SSG and the synchronized output of the AWG (a TTL square wave with the same frequency as the regular output) using a Mini-Circuits ZFM-3H mixer. Using different outputs than those of the modulation signals prevents crosstalk that can lead to an interfering signal.

All of this is controlled via a computer, which runs a Python program to control the probe laser and read out the lock-in amplifier data.

\begin{table}[h!]
    \centering
    \caption{\textbf{The experimental parameters of the components used in the setup.}}
    \begin{tabular}{c|c|c|c}
        \textbf{Parameter} & \textbf{Value} & \textbf{Unit} & \textbf{Description}\\
        \hline
        $V_{\pi, \mathrm{probe}}$ & 4.94 & V & $V_\pi$ of the probe modulator\\
        $V_{\pi, \mathrm{pump}}$ & 4.94 & V & $V_\pi$ of the pump modulator\\
        $P_{\mathrm{mod, probe}}$ & 16.0 & dBm & RF power sent to the probe modulator\\
        $P_{\mathrm{mod, pump}}$ & 6.0 & dBm & RF power sento to the pump modulator\\
        $r_{\mathrm{pd}}$ & 0.8 & A/W & Photodiode sensitivity
    \end{tabular}
    \label{tab:meas_gen}
\end{table}

\begin{table}
    \centering
    \caption{\textbf{The experimental parameters of the measurement of the polymer cladded sample.}}
    \begin{tabular}{c|c|c|c}
        \textbf{Parameter} & \textbf{Value} & \textbf{Unit} & \textbf{Description}\\
        \hline
        $P_{\mathrm{probe}}$ & 21.2 & dBm & Probe optical power after amplification\\
        $P_{\mathrm{pump}}$ & 27 & dBm & Pump optical power after amplification\\
        $\alpha_{\mathrm{wg}}$ & 1.0 & dB/cm & Optical waveguide loss\\
        $\alpha_{\mathrm{c}} $ & 8.5 & dB/facet & Coupling loss per facet, including fiber components
    \end{tabular}
    \label{tab:meas_NoSiO}
\end{table}

\begin{table}
    \centering
    \caption{\textbf{The experimental parameters of the silicon oxide and polymer cladded sample.}}
    \begin{tabular}{c|c|c|c}
        \textbf{Parameter} & \textbf{Value} & \textbf{Unit} & \textbf{Description}\\
        \hline
        $P_{\mathrm{probe}}$ & 15.6 & dBm & Probe optical power after amplification\\
        $P_{\mathrm{pump}}$ & 23 & dBm & Pump optical power after amplification\\
        $\alpha_{\mathrm{wg}}$ & .59 & dB/cm & Optical waveguide loss\\
        $\alpha_{\mathrm{c}}$ & 5.8 & dB/facet & Coupling loss per facet, including fiber components
    \end{tabular}
    \label{tab:meas_SiO}
\end{table}

From these lock-in measurements, the Brillouin gain of a waveguide can be determined as:

\begin{equation}\label{eq:SBS_gain}
    G_\mathrm{B} = e^{P_\mathrm{pump}L_\mathrm{eff}g_\mathrm{B}},
\end{equation}

where the effective length of a waveguide is calculated using:
\begin{equation}
    L_{\rm{eff}} = \frac{1-e^{-\alpha L}}{\alpha}.
\end{equation}

Here, $\alpha$ is the propagation loss, and $L$ is the actual waveguide length. By using \eqref{eq:SBS_gain}, and taking the small signal approximation ($e^x = 1+x$) we can calculate the gain coefficient using:
\begin{equation}
g_{\rm{B}, \rm{chip}} = \frac{V_{\rm{chip}}}{V_{\rm{fiber}}}\frac{g_{\rm{B}, \rm{fiber}}L_{\rm{eff, fiber}}P_{\rm{pump,fiber}}}{L_{\rm{eff, chip}}P_{\rm{pump,chip}}}.
\end{equation}

Here $V$ denotes the signal voltage measured by the lock-in amplifier, the subscripts \rm{fiber} and \rm{chip} refer to the properties of the fiber and chip used in this experiment.

The measurements for the RF filter and SBS gain using a VNA were performed using an IQ modulator driven by a VNA. In this setup, as illustrated in figure \ref{fig:fig3}~(a), the probe laser (NewFocus TLB-6728-P) was modulated using an IQ modulator based on a dual parallel Mach-Zehnder modulator (DPMZM, Covega LN86S-FC) with the RF inputs from a hybrid coupler. For the SBS gain measurement, the DPMZM was biased to have a lower frequency single-sideband (SSB) modulation output where the sideband is used to sense the SBS amplification. The complete probe signal is amplified to 33 dBm using an erbium doped fibre amplifier (EDFA, Amonics AEDFA-37-R), which gave a sideband of 8.4 dBm, whilst the pump laser (Avanex LMI1915, DFB) at 1550 nm is amplified with an EDFA (Amonics AEDFA-33-B) to 33.6 dBm.

The pump and probe are coupled to a 15 cm spiral of the cladding engineered sample with a total insertion loss from the EDFA to the power meters of 24 dB. The transmitted probe is detected using a photodiode (Optilab PD-23-C-DC), for the VNA based gain measurement the filter indicated in the figure was not present. 

For the Brillouin based RF filter, the IQ modulator was biased to be near phase modulation with a slight imbalance between lower and upper sideband. In addition, the pump laser was replaced with a laser (Agere D2525P20, DFB) at 1561 nm, and the transmitted signal is filtered before detection to get rid of the majority of the reflected pump using a tunable filter (TeraXion TFN-1561.000-N2-IL6.5-20-C1P-C2). The sample and other settings remained the same between the VNA based gain and RF filter measurements.

\section*{Supplementary note B: SBS RF photonic notch filter}

\begin{figure}[h!]
    \centering
    \includegraphics[width=0.35\textwidth]{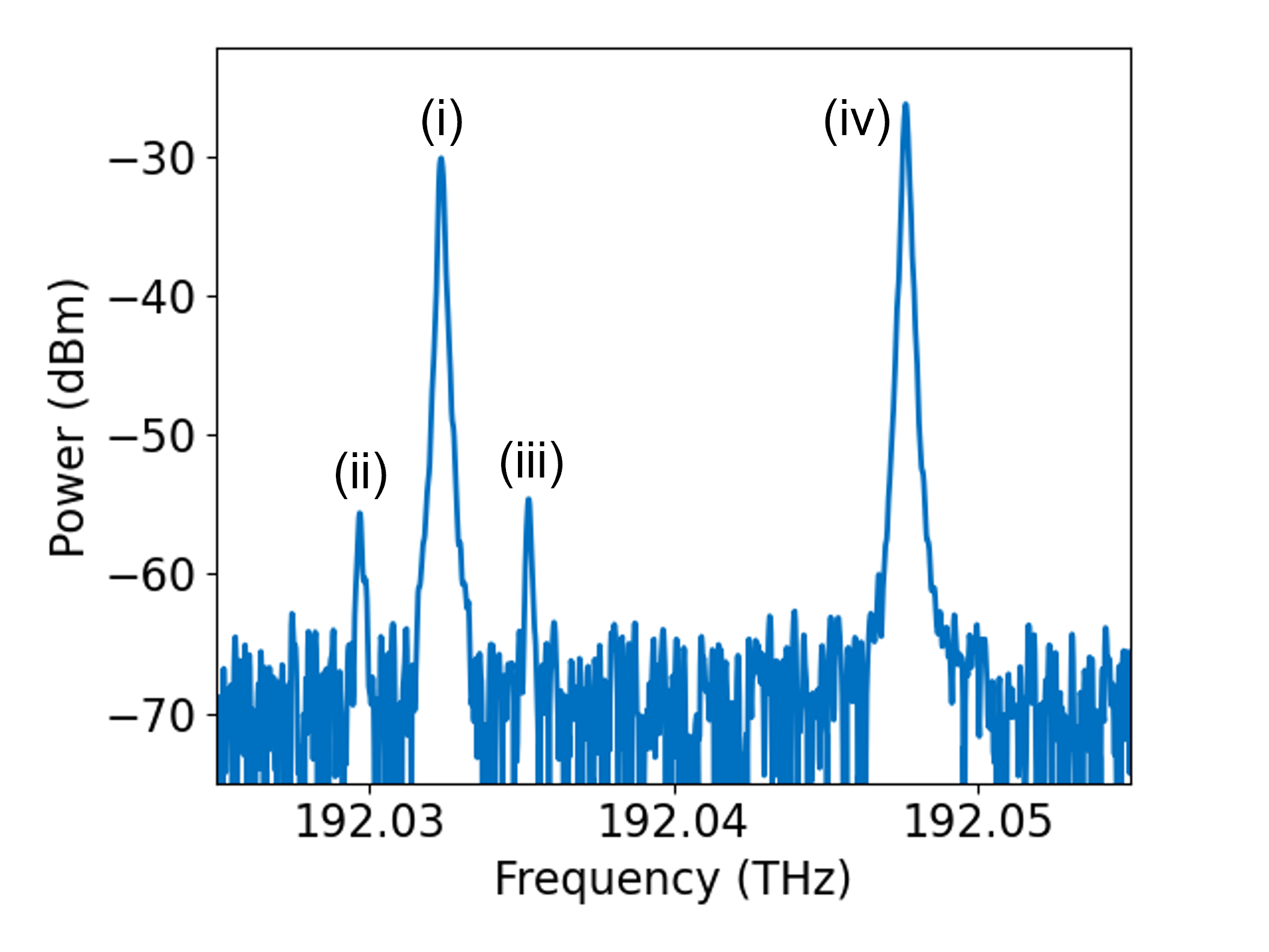}
    \caption{\textbf{OSA recorded snapshot of the optical spectrum before the photodiode in an SBS RF photonic notch filter measurement, in the Stokes configuration.} Indicated are (i) the probe carrier, (ii, iii) the probe LSB and USB, and (iv) the pump laser.}
    \label{fig:filter_OSA_setup}
\end{figure}

To demonstrate the utility of large SBS gain in silicon nitride circuits, we demonstrate an SBS-based RF photonic notch filter. To set up this filter, a probe laser and DPMZM are used to obtain DSB modulation, set close to perfect phase modulation by minimising the power received at the photodiode. Next, the upper sideband of the probe carrier (i.e., the probe) is attenuated slightly by adjusting the DPMZM biasing. By placing the pump laser at an appropriate frequency detuning from the probe carrier, the on-chip SBS gain/loss can give a frequency-dependent enhancement of the signal cancellation after photodetection. This setup is shown in Fig.~\ref{fig:filter_OSA_setup}. 

After setting up the modulation, first, the RF signal is minimized with the pump off. At this point, close to perfect phase modulation is obtained. Now, turning on the pump, a clear signal due to Brillouin scattering in the fibres is observed. The relative power of the USB and LSB is tuned through the DPMZM biasing voltages, until a notch is seen at the Brillouin shift as determined by the VNA gain measurement. The same filtering can be obtained by placing the pump line red detuned from the probe LSB, in which case the cancellation is due to loss through anti-Stokes scattering.

\begin{figure}[h!]
    \centering
    \includegraphics[width=\textwidth]{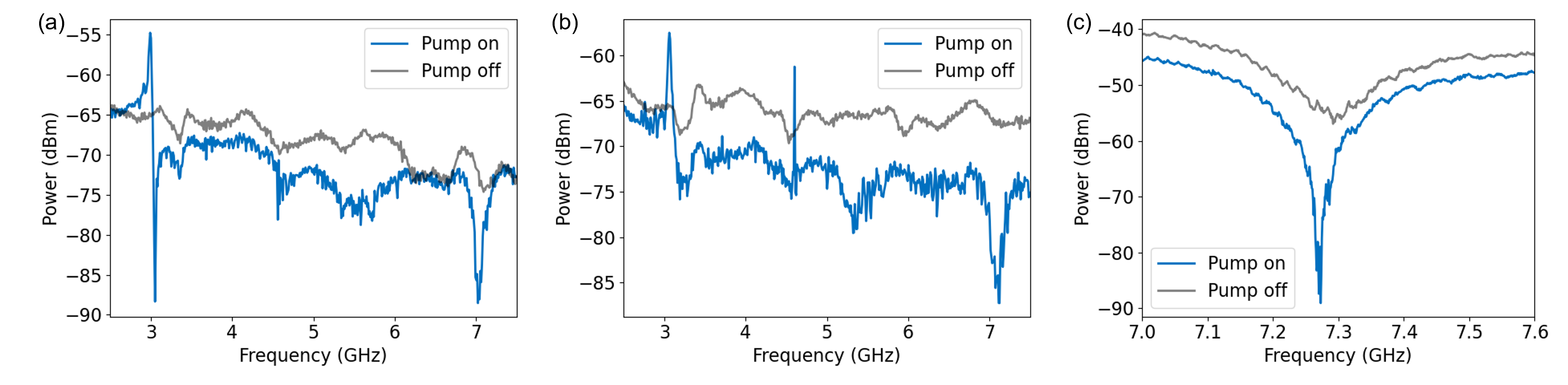}
    \caption{\textbf{The SBS RF photonic notch filter measurements.} Measured RF spectrum with pump on and off, for the filter in (a) the Stokes configuration, (b) the anti-Stokes configuration and (c) the Stokes configuration in a short 1~cm waveguide}
    \label{fig:filter_measurements}
\end{figure}

Filters using both  Stokes and anti-Stokes scattering are measured and shown in Fig.~\ref{fig:filter_measurements} (a) and (b), respectively. The notch response is confirmed to be due to the SBS gain at 6.8~GHz by inspection of the frequency spacing with response from the fibre SBS gain peak. The rejection of this filter is about 15~dB, which is limited by the small ($\sim$0.2 dB) on-chip SBS gain measured in the VNA setup, as well as modulator imperfections. The out-of-band power level is also very low, because of the use of near phase modulation. However, with improved gain and other platform-specific advantages, such as low propagation loss and erbium-doping, this result shows promise for high-performance integrated photonic RF filtering.

This method of generating a filter with very small signal gain can be done in even smaller waveguides where in Fig.~\ref{fig:filter_measurements}~(c) a Stokes based filter is performed in a 1~cm waveguide, where the same pump and probe power was utilized. In this measurement the SBS added an increased rejection of over 20 dB, where the base line measurement already rejects for 10 dB due to the frequency dependence of the modulation state of the DPMZM being significant this close to the pure phase modulated state.

\section*{Supplementary note C: Details on the simulations}

In this work to simulate the opto-acoustic interaction we used the COMSOL Multiphysics model we previously developed for our multilayer silicon nitride waveguides \cite{Botter_SciAdv_2022}. In this model, we first simulate the optical mode of the pump and probe. We then use this to calculate the electrostrictive stress tensor and optical forces induced by the electrostriction effect. We then use these as a source in our acoustic model. We model the detuning between pump and probe by changing the frequency at which the force oscillates. Finally we calculate the Brillouin gain coefficient based on the overlap between the optical forces $\mathbf{\tilde{f}}_\mathrm{n}$ and the acoustic response $\mathbf{\dot{\tilde{u}}}_\mathrm{n}$ \cite{Rakich_PRX_2012}:

\begin{equation}
    g_\mathrm{B} = 2 \cdot \left( \frac{\omega_\mathrm{s}}{\Omega} \right) \int_{\mathrm{wg}} \mathrm{Re} \left[ \mathbf{\tilde{f}}_\mathrm{n} \left( \mathbf{x}, \mathbf{y} \right) \cdot \mathbf{\dot{\tilde{u}}}_\mathrm{n} ^* \left(\mathbf{x}, \mathbf{y} \right) \right] dA,
\end{equation}

where $\omega_\mathrm{s}$ is the angular frequency of the Stokes wave and $\Omega$ the angular frequency of the excited acoustic mode. 
The peak value of the Brillouin gain coefficient can be simplified by \cite{Agrawal_2013}:

\begin{equation}
    g_\mathrm{B} = \frac{4\pi^2\gamma_\mathrm{e}^2\eta}{n_\mathrm{p}c\lambda_\mathrm{p}^2\rho_0v_\mathrm{A}\Gamma_\mathrm{B}A_{\mathrm{eff}}}.
\end{equation}

where $\gamma_\mathrm{e}$ is the electrostrictive constant, $\eta$ the acousto-optic overlap, $n_\mathrm{p}$ the refractive index, $\lambda_\mathrm{p}$ the pump wavelength, $\rho_0$ the density, $v_\mathrm{a}$ the speed of sound, $\Gamma_\mathrm{B}$ the Brillouin linewidth, and $A_{\mathrm{eff}}$ the effective area.

Table~\ref{tab:mat_data} shows the material parameters used in our simulation.

\begin{table}[h]
    \centering
    \caption{\textbf{Material data used in the simulations.} The properties labelled \textsuperscript{\textdagger} are speculated but of very small impact due to low interaction of said property, and the properties labelled \textsuperscript{*} are approximated based on the measured Brillouin linewidth.}
    \begin{tabular}{c|c|c|c|c|c|c|c}
        \textbf{Material} & \textbf{Refractive} & \textbf{Density} & \textbf{Young's} & \textbf{Poisson's} & \textbf{Bulk} & \textbf{Shear} & \textbf{photoelastic} \\
         & \textbf{index} &  & \textbf{Modulus} & \textbf{ratio} & \textbf{viscosity} & \textbf{viscosity} & \textbf{coefficient} \\
         &  &  (kg/m$^3$) & (GPa) &  & (×10$^{-3}$ Pa$\cdot$s) & (×10$^{-3}$ Pa$\cdot$s) & $\mathbf{p_{12}}$ \\
         \hline
        SiO$_2$ \cite{Håkansson_APL_2019,Smith_OL_2016}& 1.49 & 2203 & 73.1 & 0.17 & 1.6 & 0.16 & 0.27\\
        Si$_3$N$_4$ \cite{Gundavarapu_NatPhot_2018,Gyger_PRL_2020} & 2.0 & 3020 & 201 & 0.23 & 1.6\textsuperscript{\textdagger} & 0.16\textsuperscript{\textdagger} & 0.047\\
        TeO$_2$ \cite{Yano_JApplPhys_1971} & 2.13 & 5870 & 45 & 0.25 & 1.8 & 0.18 & 0.241\\
        CYTOP \cite{Cytop_Brochure} & 1.3335 & 2030 & 1.5 & 0.42 & 18\textsuperscript{*} & 1.8\textsuperscript{*} & 0.108\textsuperscript{\textdagger}
    \end{tabular}
    \label{tab:mat_data}
\end{table}

\section*{Supplementary note D: Further waveguide simulations}

The measured samples had different layer heights for the silicon nitride and tellurite. The layer heights will also affect the Brillouin response, but we expect this to be significantly smaller than the changes made to the cladding. We performed additional simulations to confirm our expectations.

\begin{table*}
\centering
    \caption{\textbf{Extended table of measurement and Simulation results of different waveguide geometries of the tellurite covered silicon nitride waveguide.} The geometry is varied by the height of the silicon nitride layer ($h_{Si_3N_4}$), the width of the silicon nitride layer ($w_{Si_3N_4}$), the height of the tellurite layer ($h_{TeO_2}$), height of the silica layer ($h_{SiO_2}$) and a CYTOP cover. For each geometry the maximum Brillouin gain coefficient (max $g_B$) with the corresponding Brillouin shift (SBS shift) is given as both simulated and measured results.}
\begin{tabular}{c|c|c|c|c|c|c|c|c|c}
\textbf{Geometry}  & $\mathbf{h_{Si_3N_4}}$ & $\mathbf{w_{Si_3N_4}}$ & $\mathbf{h_{TeO_2}}$ & $\mathbf{h_{SiO_2}}$ & \textbf{CYTOP}  & \multicolumn{2}{c}{\textbf{Simulated}} & \multicolumn{2}{c}{\textbf{Measured}}               \\
  &  (nm)  & (nm) & (nm) & (nm) &  \textbf{cover} & \textbf{max} $\mathbf{g_B}$ & \textbf{SBS shift} & \textbf{max} $\mathbf{g_B}$ & \textbf{SBS shift} \\  
  &   &   &   &   &   & (m$^{-1}$W$^{-1}$) & (GHz) & (m$^{-1}$W$^{-1}$) & (GHz) \\
                           \hline 
Polymer cladding                & 100 & 1600 & 354 & 0 & yes & 6.5 & 7.47 & 4.5 & 8.2 \\
Thin silica cladding            & 200 & 1600 & 299 & 46 & yes & 16.4 & 8.58 & 8.5 & 8.9 \\
Full silica cladding            & 100 & 1600 & 354 & full & no & 42.5 & 7.96 & - & - \\
Polymer cladding supplement     & 200 & 1600 & 299 & 0 & yes & 5.7 & 7.85 & - & - \\
Thin silica cladding supplement & 100 & 1600 & 354 & 46 & yes & 14.5 & 7.93 & - & - \\
Optimized thin silica cladding  & 370 & 2520 & 590 & 100 & yes & 67.1 & 8.203 & - & - \\
Optimized full silica cladding  & 110 & 4020 & 270 & full & no & 154.8 & 6.335 & - & - 
\end{tabular}
\label{tab:gain_extended}
\end{table*}

\subsection*{200 nm silicon nitride tellurite covered waveguide without silicon oxide layer}

The geometry of the waveguide from Fig.~\ref{fig:fig4}, but without the silcon oxide cladding can  be seen in Fig.~\ref{fig:figS6}~(a). Here, the waveguide consists of a 200~nm high silicon nitride layer, covered with a 299~nm tellurite layer, only cladded in polymer. Fig.~\ref{fig:figS6} (b) and (c) show the optical and acoustic responses respectively. These look similar to those in Fig.~\ref{fig:fig2} (b) and (c), with the acoustic waves leaking into the polymer. Fig.~\ref{fig:figS6}~(g) shows the resulting Brillouin gain spectrum. Here the gain peaks at 5.7~m$^{-1}$W$^{-1}$, at a frequency shift of 7.85~GHz. This is slightly lower than the waveguide with a thinner layer of silicon nitride discussed in the main text. This difference is likely due to a smaller part of the optical field being in the tellurite layer, as depicted in Fig.~\ref{fig:figS6}~(b), reducing the acousto-optic overlap.

\subsection*{100 nm silicon nitride tellurite covered waveguide with silicon oxide layer}

\begin{figure}[t!]
    \centering
    \includegraphics[width=\textwidth]{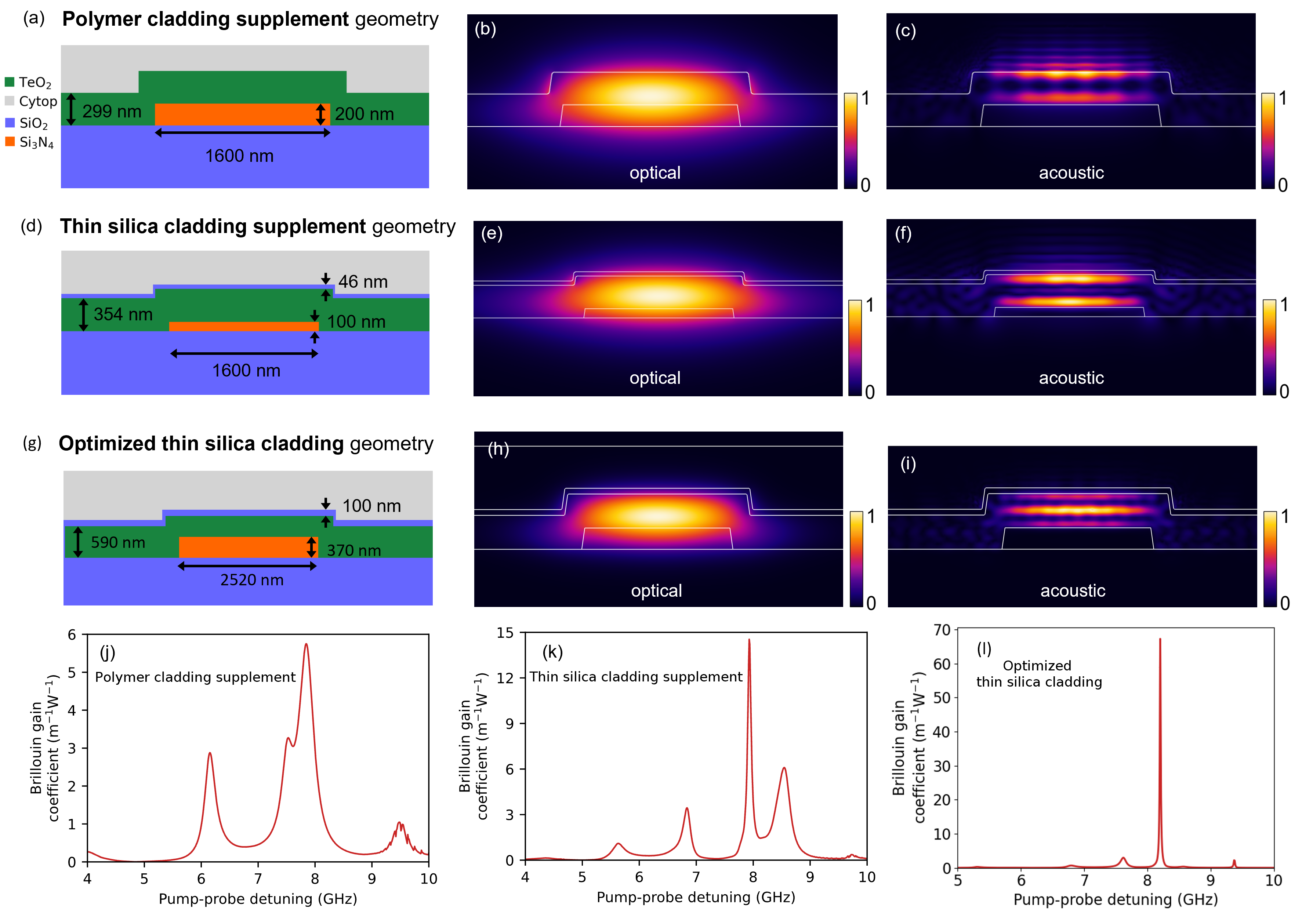}
    \caption{\textbf{The tellurite covered silicon nitride waveguide, with a (a-c,h) 200~nm high silicon nitride layer, a 299~nm tellurite layer and a polymer cladding, or with a (d-f,g) a 100~nm high silicon nitride layer, a 354~nm tellurite layer, and a cladding made of a 46~nm silicon oxide layer and polymer.} (a,b) the geometry of the waveguide, (g,h) the simulated Brillouin response. The simulation results, with (b,e) the electrical field of the optical mode, and (c,f) the displacement field of the acoustic response at the peak of the Brillouin response.}
    \label{fig:figS6}
\end{figure}

Fig.~\ref{fig:figS6}~(d) shows the geometry of a waveguide consisting of a 100~nm high silicon nitride layer, covered with a 354~nm tellurite layer, and cladded with a 46~nm layer of silicon oxide, and finally the polymer. This mirrors the waveguide depicted in Fig.~\ref{fig:fig2}~(a), but with the additional silicon oxide layer in the cladding

Fig.~\ref{fig:figS6} (e) and (f) show the optical and acoustic responses respectively, and Fig.~\ref{fig:figS6}~(h) shows the resulting Brillouin gain spectrum. The gain peak coefficient is 14.5~m$^{-1}$W$^{-1}$, corresponding to a frequency shift of 7.93~GHz, similar to that seen in Fig.~\ref{fig:fig2}~(j). Here, the gain coefficient is higher than in the main body of the paper, because, as shown in Fig.~\ref{fig:figS6}~(e), a larger portion of the optical field is in the tellurite, increasing the acousto-optic overlap, and thus the Brillouin gain.

From these simulations we can conclude that the higher gain simulated and measured in the second sample is primarily due to the additional silicon oxide layer, and not significantly due to the difference in layer thickness.

\section*{Supplementary note E: Genetic algorithm optimization}

\subsection*{Tellurite covered silicon nitride waveguide with silicon oxide layer and CYTOP polymer claddding}

To determine how much of gain is achievable with current fabrication limitations, a genetic optimization of the Brillouin gain is carried out for the tellurite covered silicon nitride waveguide with a 100~nm layer of silicon oxide and CYTOP polymer cladding. Details about the genetic algorithm which is used can be found in \cite{Håkansson_APL_2019}. The gain is optimized by varying the geometric parameters of the waveguide cross-section. A fixed value of 100~nm is assumed for the silicon oxide cladding layer. The width and height of the silicon nitride layer and the height of the tellurite and CYTOP layers are allowed to vary. The search space is limited to $w_{\mathrm{Si_3N_4}} \in$ [800, 3200]~nm, $h_{\mathrm{Si_3N_4}} \in$ [100, 400]~nm, $h_{\mathrm{TeO_2}} \in$ [150, 600]~nm and $h_{\mathrm{CYTOP}} \in$ [800, 2400]~nm. The maximum thickness of silicon nitride is chosen as the limit for crack-free LPCVD \cite{Grootes_OE_2022}. The CYTOP layer thickness was found not to affect the gain significantly beyond 800~nm, so this parameter could have been excluded.

\begin{figure}[h!] 
    \centering
    \includegraphics[width=0.9\textwidth]{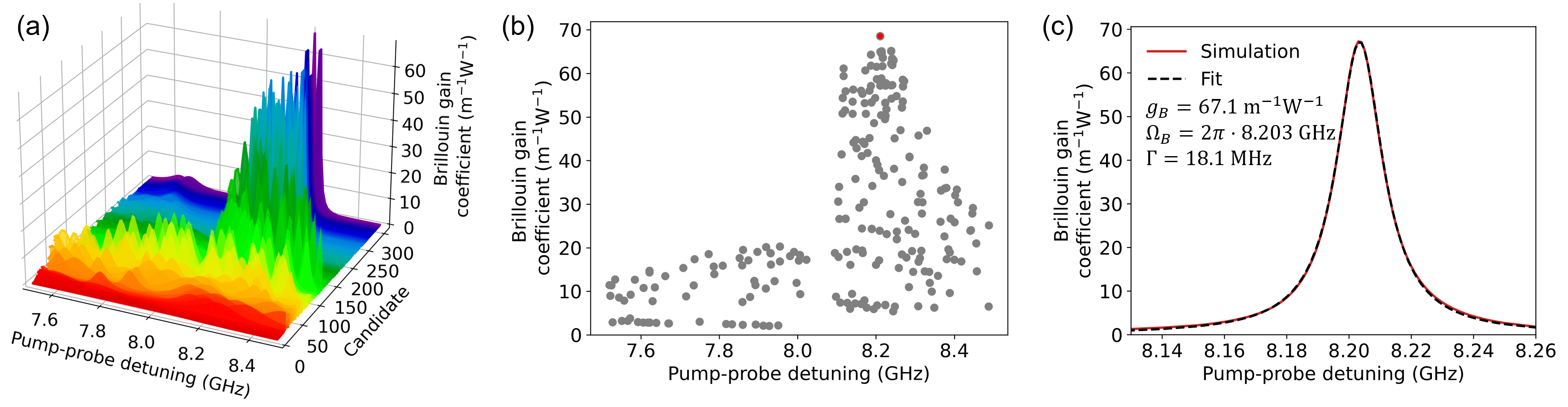}
    \caption{\textbf{Genetic optimization results for the tellurite covered silicon nitride waveguide with a silicon oxide layer and CYTOP polymer cladding.} (a) Simulated SBS spectra for each candidate, denoting a separate geometry, sorted from low to high peak gain. (b) Simulated peak SBS gain. The optimized geometry, Fig.~\ref{fig:fig4} (a-c) in the main text, is indicated with a red marker. (c) Lorentzian fit to the gain SBS gain peak of the optimized structure.}
    \label{fig:S4}
\end{figure}

Fig.~\ref{fig:S4} (a) shows the simulated SBS gain spectra for all candidates geometries in this optimization, sorted by peak gain. Here we see a convergence to single, well-defined peaks near a pump-probe detuning (acoustic frequency) of 8.2 GHz. For clarity, the peak gain values and their corresponding frequency shift are plotted in Fig.~\ref{fig:S4} (b). The maximal gain can be obtained for acoustic frequencies near 8.2 GHz, approaching 70~m$^{-1}$W$^{-1}$. The optimized result is indicated with a red marker.

The optimized geometry obtained from the optimization is shown in Fig.~\ref{fig:figS6} (g). The silicon nitride strip width and thickness are 2520~nm and 370~nm, and the tellurite thickness is 590~nm. The CYTOP layer is 1100~nm thick, although this has little influence on the gain beyond 800~nm, due to improved vertical acoustic confinement. Fig.~\ref{fig:figS6} (l) the Brillouin gain spectrum and Table~\ref{tab:gain_extended} compares the peak gain. 

The geometry of this waveguide is considerably larger than the previously considered waveguides, approaching the limits set on silicon nitride and tellurite thickness. From fitting to a Lorentzian line shape, Fig.~\ref{fig:S4} (c), the maximum gain is 67.1 m$^{-1}$W$^{-1}$ at 8.203 GHz. The linewidth of the gain peak is $\Gamma = 18.1$~MHz. Namely, a slightly lower spatial resolution was used in the optimization to reduce the computation time. Fig.~\ref{fig:figS6} (h) and (i) show the optical mode and acoustic response at the gain maximum, respectively. 

\begin{figure}[h!] 
    \centering
    \includegraphics[width=\textwidth]{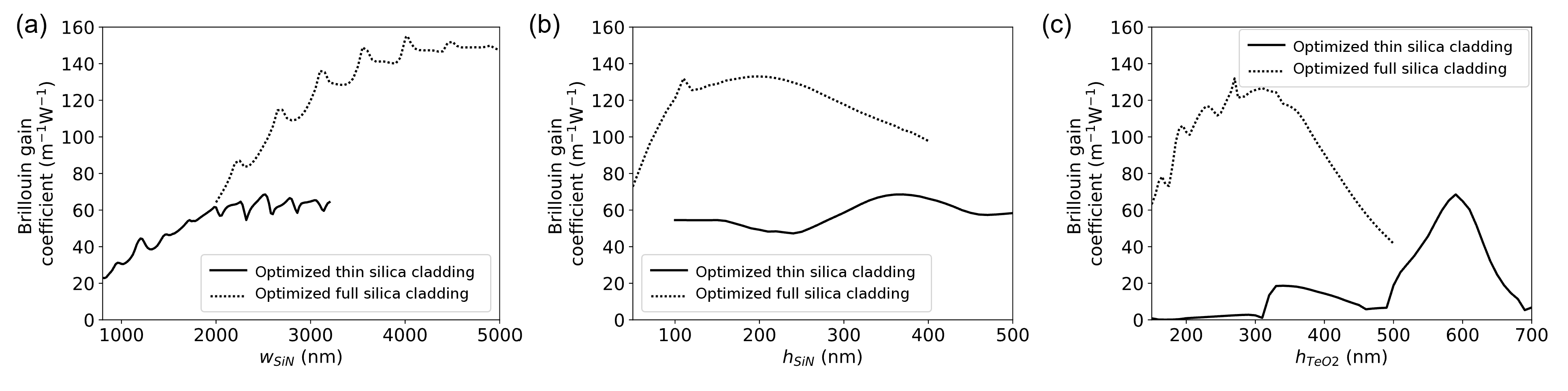}
    \caption{\textbf{Brillouin gain for varying silicon nitride width (a) and thickness (b) around the optimized geometry for the tellurite covered silicon nitride waveguide with a 100~nm silicon oxide layer and CYTOP cladding and the optimized geometry for the tellurite covered silicon nitride waveguides with full silicon oxide cladding.}}
    \label{fig:S6}
\end{figure}

To get more physical insight into the SBS dynamics in this waveguide, the geometric parameters of the optimized structure are varied around their optimal values. Fig.~\ref{fig:S6} shows the SBS gain for variations in the silicon nitride strip width (a) and thickness (b), and tellurite layer thickness (c). The SBS gain initially increases with width, decreasing beyond the optimum at 2520~nm. This is understood as a balance of increase in overlap between the radiating acoustic fields and the confined optical mode, and increasing optical mode area at large widths. Additionally, there are oscillations on top of this general trend. These are due to an increase/decrease in radiation from the waveguide core into the tellurite slab, depending on the interference pattern of the acoustic field at the edges of the waveguide core. With variation in silicon nitride height, there is not as much change in the peak gain. By varying the tellurite layer thickness, the most drastic change in gain is observed. This is because the phonons are mostly trapped in this layer. The vertical thickness determines which, if any, phonon modes are well-guided inside the core. At lower thicknesses, phonon modes with one or two lobes, instead of the three lobes in Fig.~\ref{fig:figS6} (i), can be excited. 

\subsection*{Tellurite covered silicon nitride waveguide with full silicon oxide claddding}

In order to fully explore the potential of the tellurite covered silicon nitride platform, a genetic optimization of the Brillouin gain in the silicon oxide cladded waveguide is performed. The same limits for geometric parameters of silicon nitride and tellurite are used. The silicon oxide thickness is limited to $h_{\mathrm{SiO_2}} \in$ [800, 3200]. As before, though, the cladding thickness beyond 800~nm was found to have little influence on the gain.

Because of the hardness and low acoustic loss of silica, the acoustic behaviour in this waveguide is quite different. As opposed to the CYTOP cladded waveguide, the acoustic waves dissipating into the tellurite layer are not quickly damped. At the gain peak, this can lead to unphysical results (e.g. gain depending on the width of the simulation domain), because of reflections from the PMLs which are used to absorb radiating acoustic fields. The acoustic wavevector can be close to parallel with the optical axis near a Brillouin resonance, and PMLs have significantly higher reflectance near parallel incidence. This problem was solved with an adiabatic increase in viscosity of the tellurite layer, increasing the acoustic loss. This resolved the unphysical results by damping the oscillations radiating away from the core. However, this siginificantly increased the computation time, hence the low number of candidate geometries as stated in the main text.

\begin{figure}[h!]
    \centering
    \includegraphics[width=0.67\textwidth]{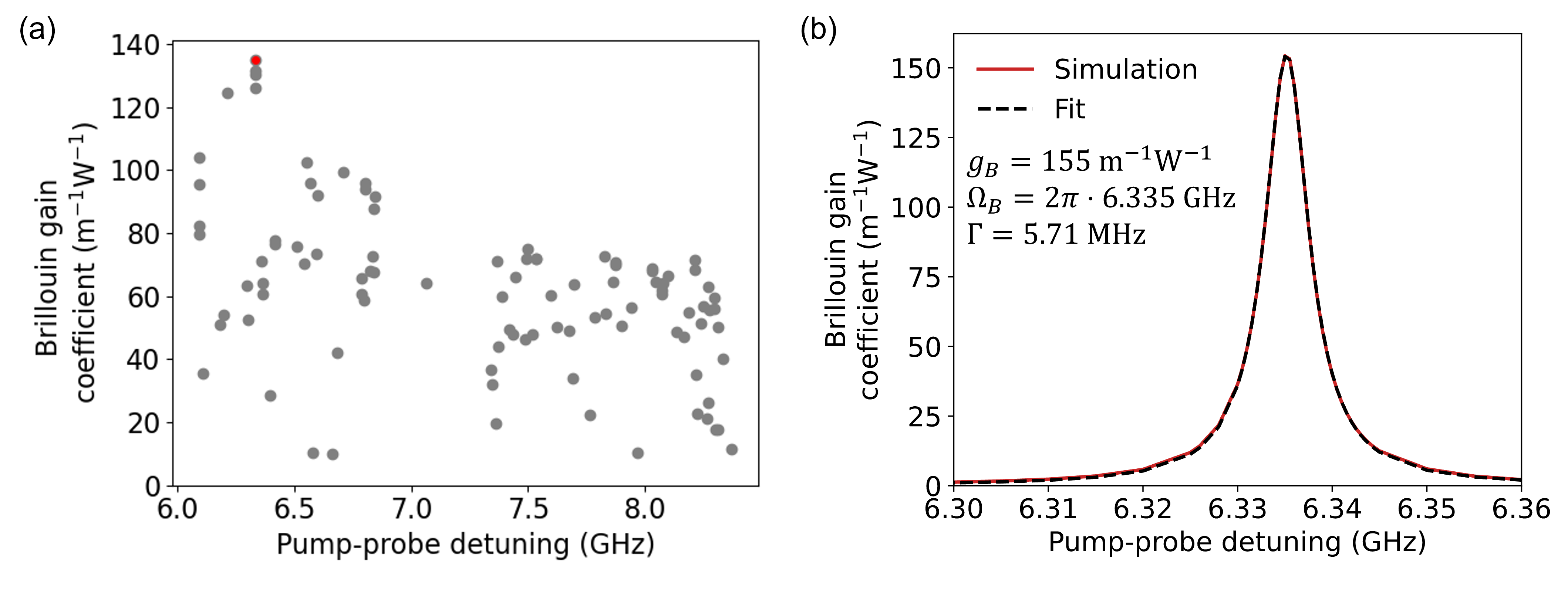}
    \caption{\textbf{Preliminary genetic optimization run of the tellurite covered silicon nitride waveguide with a full silica cladding.} (a) Simulated peak SBS gain vs. frequency shift. The result with the highest gain within the optimization is indicated with a red marker. (b) Lorentzian fit to the Brillouin gain peak of the highest gain geometry in Fig.~\ref{fig:fig4}~(g).}
    \label{fig:S10}
\end{figure}

The Brillouin gain vs. pump-probe detuning for each candidate in this preliminary optimization are shown in Fig.~\ref{fig:S10} (a). The red marker again indicates the optimized geometry. We find a peak gain of $g_B = 132$~m$^{-1}$W$^{-1}$ at a frequency shift of 6.332~GHz. The frequency dependence of the maximum gain appears less strong as in CYTOP covered waveguides.

However, in doing sweeps over the geometric parameters around the current optimum, it was found that higher gain is achieved outside the optimization window. Namely, a silicon nitride width of 4020~nm results in $g_B \approx 155$~m$^{-1}$W$^{-1}$. The 4020~nm wide waveguide is shown in Fig.~\ref{fig:fig4} (a) in the main text. Its fundamental optical mode and acoustic response at the Brillouin gain peak are shown in Fig.~\ref{fig:fig4} (b) and (c). Again, a Lorentzian fit is done to the gain peak, which is shown in Fig.~\ref{fig:S10} (b). This shows the peak gain $g_B \approx 155$~m$^{-1}$W$^{-1}$ at an acoustic frequency of 6.335~GHz. The width of the gain peak is even narrower than in the optimized geometry with a CYTOP cladding, namely $\Gamma = 5.71$~MHz.

The parameter sweeps were around the optimum geometry are shown in Fig.~\ref{fig:S6}. Fig.~\ref{fig:S6} (a) and (b) show the variation of the Brillouin gain with the silicon nitride strip width and height. As can be seen, the value of 132~m$^{-1}$W$^{-1}$ determined by the genetic algorithm so far is not a global optimum. For the tellurite width, the genetic algorithm has converged to an optimum layer thickness, though this will change for different geometries of the silicon nitride strip. Notably, compared to CYTOP cladded waveguides, the optimum is found at lower silicon nitride and tellurite thicknesses.

\end{document}